\documentclass[RNAAS]{aastex62}
\begin{document}

\title{A shortcut to calculate SPAM limb-darkening coefficients}

\correspondingauthor{Giuseppe Morello}
\email{gmorello@iac.es}

\author[0000-0002-4262-5661]{Giuseppe Morello}
\affiliation{Instituto de Astrof\'isica de Canarias (IAC), 38205 La Laguna, Tenerife, Spain}
\affiliation{Departamento de Astrof\'isica, Universidad de La Laguna (ULL), 38206, La Laguna, Tenerife, Spain}
\affiliation{INAF- Palermo Astronomical Observatory, Piazza del Parlamento, 1, 90134 Palermo, Italy}

\author{Andrea Chiavassa}
\affiliation{Universit\'e C\^ote d'Azur, Observatoire de la C\^ote d'Azur, CNRS, Lagrange, CS 34229, Nice,  France}


\begin{abstract}
We release a new grid of stellar limb-darkening coefficients (LDCs, using the quadratic, power-2 and claret-4 laws) and intensity profiles for the Kepler, U, B, V and R passbands, based on \texttt{STAGGER} model atmospheres. The data can be downloaded from Zenodo \doi{doi:10.5281/zenodo.5593162}.  We compare the newly-released LDCs, computed by \texttt{ExoTETHyS}, with previously published values, based on the same atmospheric models using a so-called ``SPAM'' procedure. The SPAM method relies on synthetic light curves in order to compute the LDCs that best represent the photometry of exoplanetary transits. We confirm that \texttt{ExoTETHyS} achieves the same objective with a much simpler algorithm.
\end{abstract}

\section{Introduction} 
Most codes that calculate transiting exoplanet light curves use limb darkening coefficients (LDCs) to approximate the radially-decreasing stellar intensity profile \citep{kreidberg2015,parviainen2015,agol2020,morvan2021}. Theoretical LDCs can be obtained by fitting a parametric law to numerical evaluations of the intensity across the stellar disk. A long list of limb darkening laws have been implemented, for which we refer to notable literature reviews \citep{kipping2013,espinoza2016,agol2020}. For a given law, the choice of the cost function and the sampling of the intensity profile affect the estimated LDCs. What are the best practices for determining theoretical LDCs has been the subject of a long debate (see \citealp{morello2021} and references therein).

\section{The SPAM method}
\cite{howarth2011} pointed out that what is \textit{best} generally depends on the purpose.
He produced \textit{synthetic-photometry} light curves using \textit{atmosphere-model} intensity profiles for a number of well-known systems, then solved for the geometric parameters and LDCs. The resulting SPAM LDCs are the most direct reference with which to compare the empirical LDCs obtained by fitting the transit light curves. Note that, by construction, the SPAM LDCs depend on the geometric parameters in addition to the stellar ones.

\cite{morello2017} continued this experiment by exploring the behavior of various limb darkening laws as a function of stellar type and wavelengths. Additionally, they included the effect of spherical geometry to generate the synthetic light curves. It appeared that the intensity profiles obtained with SPAM LDCs, there referred to as ``empirical'' LDCs, provide a better match to the numerical intensities in the inner part of the stellar disk, while the discrepancy increases towards the edge. Theoretical LDCs obtained with a simple least squares fit distribute the error more evenly, forming two or more intervals in which the intensities are either overestimated or underestimated.

\section{The \texttt{ExoTETHyS} algorithm}
\cite{morello2020} compared an exhaustive list of methods for calculating LDCs from numerical intensity values, considering differently weighted and unweighted least squares fits, resampling of the intensities via interpolation, and truncation near the limb. They considered several figures of merit based on the results of synthetic light curve fits with fixed LDCs. These figures were (1) the amplitude of the light curve residuals, (2) the biases in the recovered transit depth and other geometric parameters, and (3) their spectral variation over the 0.25-10 $\mu$m range. All criteria converged to the same best procedure, there called ``weighted-$r$ QS''. This method consists of a weighted least squares fit, where the weight associated with a given intensity value is half the radial separation between the two neighboring points. A cutoff is applied to the spherical intensity profiles.
The optimal fitting method has been confirmed by many more tests than those reported by \cite{morello2020}, including different limb darkening laws and stellar types.

The \texttt{ExoTETHyS}\footnote{\url{https://github.com/ucl-exoplanets/ExoTETHyS}} package includes a stellar LDCs calculator that implements the optimal fitting method, ensuring a precision of few parts per million for transit modeling \citep{morello2020joss}.
We note that \texttt{ExoTETHyS} should return the equivalent of SPAM LDCs, as the first figure of merit for the adopted fitting method was to provide the best match to the synthetic light curves.

\section{Independent confirmation of \texttt{ExoTETHyS}}
\cite{maxted2018} compiled a table of stellar LDCs based on synthetic stellar spectra\footnote{Publicly available on Pollux database: \url{http://npollux.lupm.univ-montp2.fr}} \citep{chiavassa2018} computed for the stellar convection simulations of the \texttt{STAGGER}-grid \citep{magic2013}. They adopted the power-2 limb darkening law \citep{hestroffer1997},
\begin{equation}
\frac{I(\mu)}{I(1)} = 1 - c \left ( 1 - \mu^{\alpha} \right ) ,
\end{equation}
that outperforms other two-coefficient laws, especially for the M-dwarf models \citep{morello2017}.
Here $\mu=\cos{\theta}=\sqrt{1-r^2}$, where $\theta$ is the angle between the surface normal and the line of sight, and $r$ is the radial distance from center of the stellar disk in units of its radius.
In addition to the coefficients $c$ and $\alpha$, \cite{maxted2018} has tabulated the transformed coefficients,
\begin{eqnarray}
&h_1& = I \left ( \frac{1}{2} \right ) = 1 - c \left ( 1 - 2^{-\alpha} \right ) \text{ and}\\
&h_2& = I \left ( \frac{1}{2} \right ) - I \left ( 0 \right ) = c 2^{-\alpha} ,
\end{eqnarray}
which were found to be less strongly correlated.
The grid of LDCs published by \cite{maxted2018} is the only one that, according to our knowledge, is based on a SPAM-like procedure. The LDCs were derived for a standard system configuration, neglecting their dependence on the geometric parameters.

\begin{figure}[h!]
\begin{center}
\includegraphics[width=0.95\textwidth]{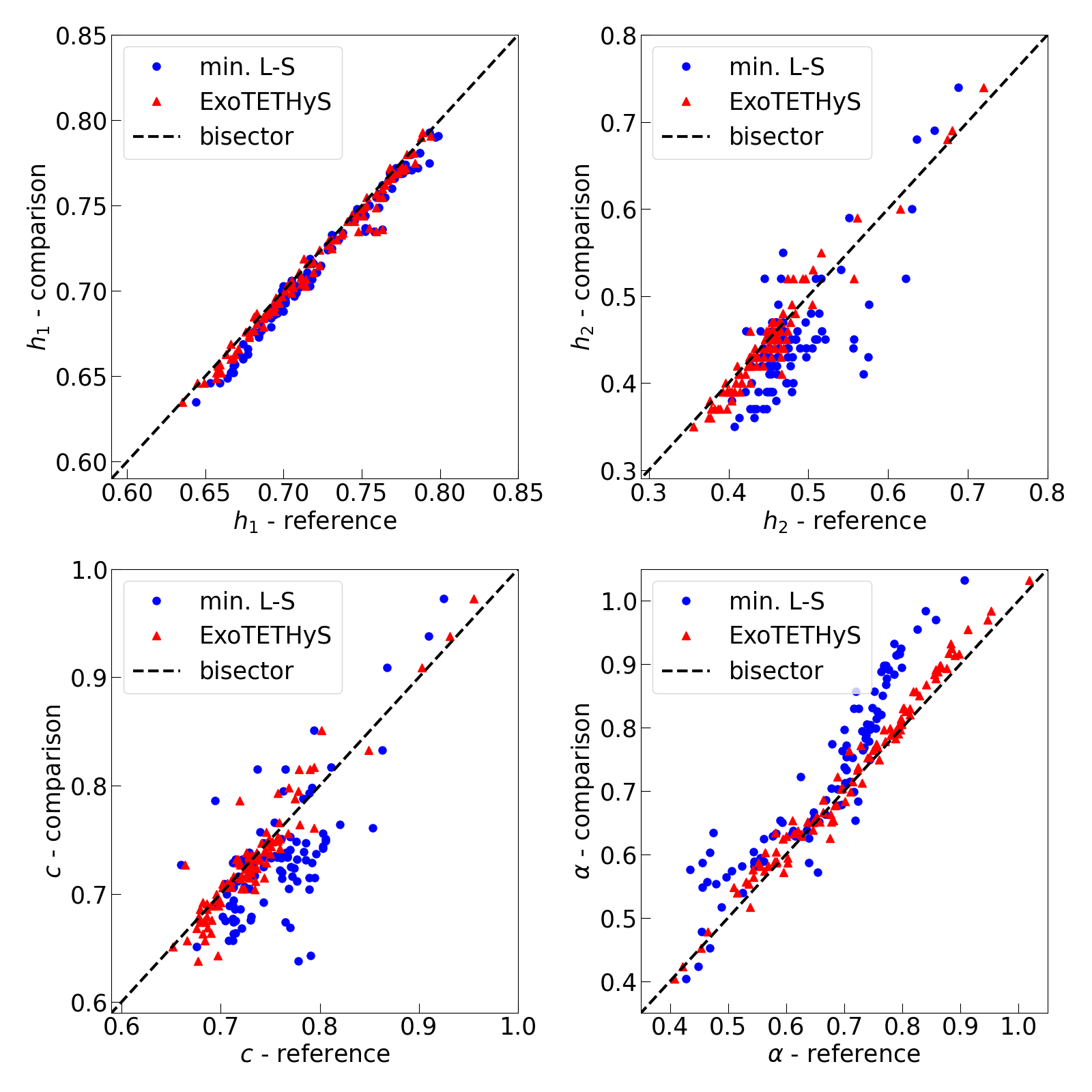}
\caption{LDCs obtained by \texttt{ExoTETHyS} (red triangles) and unweighted least squares fit (blue circles) vs. LDCs tabulated by \cite{maxted2018} for the Kepler passband.
\label{fig:ldcs_comparison_kepler}}
\end{center}
\end{figure}

The \texttt{ExoTETHyS} package includes a precalculated grid of stellar intensity spectra obtained from a subset\footnote{The new larger database will be incorporated by the end of 2021.} of the same \texttt{STAGGER} model atmospheres \citep{magic2015,chiavassa2018}. We performed grid calculations to compare the LDCs returned by \texttt{ExoTETHyS} with the SPAM-like ones tabulated by \cite{maxted2018}. To appreciate the quality of the agreement between the LDCs obtained with both methods, we calculated analogous sets of LDCs using an unweighted least squares fit. Figure \ref{fig:ldcs_comparison_kepler} shows the results obtained for the Kepler passband. The median absolute discrepancies between the \texttt{ExoTETHyS} and SPAM-like LDCs are $7.5 \times 10^{-3}$, $1.8 \times 10^{-2}$, $3.0 \times 10^{-3}$ and $1.05 \times 10^{-2}$ for $c$, $\alpha$, $h_1$ and $h_2$, respectively.
The results for other passbands are similar within a factor 0.5--2.
The discrepancies between the unweighted least squares and SPAM-like LDCs are typically larger by a factor 2--4 for $c$, $\alpha$ and $h_2$ and 1--2 for $h_1$.
We provide data files (beyond those shown in the figure) and scripts to ensure the reproducibility of our results by any user; these reproducibility materials have been deposited on Zenodo: \doi{doi:10.5281/zenodo.5593162}.

Finally, we investigate whether the difference between the LDCs obtained in this work and those previously tabulated are solely due to the fitting method. We note that the normalized intensity profiles returned by \texttt{ExoTETHyS} are similar, but not identical, to those published online. Despite both calculations rely on the same stellar atmosphere models, numerical differences must have propagated through the process to get passband-integrated normalized intensity profiles (e.g., using different spectral synthesis tools, interpolation and/or passband reference files). For the Kepler passband, the median absolute discrepancy between the intensities at the limb is 0.68\%, but it can be up to 6\% for some stellar models. The small differences between \texttt{ExoTETHyS} and tabulated LDCs could be dominated by the differences between the underlying intensity profiles, hence suggesting an even better agreement between the \texttt{ExoTETHyS} and SPAM methods.

\section{Conclusions}
We computed a grid of LDCs and intensity profiles for Kepler, U, B, V and R passbands by using \texttt{ExoTETHyS} and \texttt{STAGGER} atmospheric models. We compared our new LDCs with those reported by \cite{maxted2018} using a SPAM-like procedure that is designed to match exoplanetary transit observations. Our new LDCs are in excellent agreement with the tabulated values. These results confirm previous claims that the \texttt{ExoTETHyS} algorithm is essentially equivalent to the SPAM procedure, but without the need to compute synthetic transit light curves, hence requiring a much shorter computing time.

\acknowledgments
G. M. has received funding from the European Union's Horizon 2020 research and innovation programme under the Marie Sk\l{}odowska-Curie grant agreement No. 895525.

\bibliography{mybib.bib}

\end{document}